\documentclass[journal]{IEEEtran}
\usepackage{amsmath,amsthm, amsfonts,graphicx}
\usepackage{algorithm, algpseudocode}

\makeatletter
\newcommand{\thickhline}{
  \noalign{\ifnum0=`}\fi\hrule height 1.2pt
  \futurelet\reserved@a\@xhline
}
\makeatother

\newtheorem{prop}{Proposition}

\title{Classification Filtering}
\author{\.Ilker Bayram\\ibayram@ieee.org\\Whoop, Boston, MA, USA \thanks{This is a preprint of a paper accepted at IEEE Internatioal Workshop on Machine Learning for Signal Processing (MLSP), 2025.}}

\begin{document}

\maketitle

\begin{abstract}
We consider a streaming signal in which each sample is linked to a latent class. We assume that multiple classifiers are available, each providing class probabilities with varying degrees of accuracy. These classifiers are employed following a straightforward and fixed policy. In this setting, we consider the problem of fusing the output of the classifiers while incorporating the temporal aspect to improve classification accuracy. We propose a state-space model and develop a filter tailored for realtime execution. We demonstrate the effectiveness of the proposed filter  in an activity classification application based on inertial measurement unit (IMU) data from a wearable device.
\end{abstract}

\section{Introduction}
\label{sec:intro}

Consider a dynamic classification problem, where at each instant, a streaming signal is associated with a latent class. The true class evolves over time following an unknown transition probability. Multiple classifiers of varying strength are available but no training data. Additionally, suppose that the accuracy of classifiers are directly proportional to computational complexity. In this setting, a simple strategy is to resort to strong classifiers less often, and rely on the weak ones most of the time. The resulting challenge is to fuse outputs from classifiers of varying strengths. We present a light-weight algorithm to address this problem.

The scenario described is particularly relevant to embedded applications involving wearable devices. For example, one may wish to track a person’s activity, which naturally varies over time. In such applications, minimizing power consumption is crucial. This encourages the use of classifiers with low complexity, as they help conserve energy. To support the assumption of unavailable training data, we note that many chip manufacturers now bundle pre-trained on-chip classification algorithms, though users typically do not have access to the original training data. Consequently, any temporal smoothing must be carried out by the user.

Hidden Markov models (HMM) \cite{rab89p257, Bishop, murphy2} offer a principled approach to account for temporal dynamics in classification tasks (see also \cite{wil18} in the context of activity classification from IMUs). However, HMMs usually require access to training data and need to be trained from scratch, which is incompatible with our setting. Contemporary methods such as CNNs, RNNs, LSTMs, GANs, or transformers (see e.g., \cite{Bishop_deep} for a general treatment with extensive references, or \cite{sha21p540, zil23p355, ram21p029, dem20p816} in the context of activity recognition from IMUs), can incorporate memory or a sliding window to capture temporal structure. This can  reduce the need for additional temporal smoothing. However, even with such methods, it can be advantageous to trade  performance for lower computational cost, especially when some of that performance can be recovered through temporal smoothing. Moreover, when multiple classifiers are available, fusing their outputs naturally arises in practice. We demonstrate via numerical experiments that the accuracy of transformers can be improved with temporal smoothing.

We adopt a Bayesian framework in our approach \cite{Sarkka_2013, Bishop}. We assume that the classifiers produce class probabilities, which we treat as noisy observations of an underlying true probability distribution. We also assume that the level of noise depends on the strength of the classifier. This in turn allows us to adjust the weight given to an observation and differentiate classifiers. By specifying how the classes are expected to evolve over time, we establish a state-space formulation, and derive the corresponding update rules for the parameters being tracked.

The general Bayesian framework outlined above is, of course, well-established. However, it does not provide a turn-key solution. In most cases, the burden falls on the user to allocate computational effort toward evaluating integrals lacking closed-form expressions or  to employ particle filters capable of handling arbitrary distributions. Our contribution lies in identifying suitable assumptions and approximations that lead to an effective, lightweight solution. We emphasize that the paper does not assert temporal smoothing is universally required, or that modern methods are incapable of learning to produce time-smoothed estimates. Rather, we argue that in many real-world scenarios, the conditions under which models are trained may not align with those encountered during deployment. As a result, there is a need for a simple method to enhance the performance of black-box models. The paper offers one such method.

\subsubsection*{Outline}
Sec.~\ref{sec:dynamical_model} describes a dynamical model for observations along with a latent distribution (or, state) we would like to track.
This is followed by a derivation of update equations for the parameters of the latent distribution in Sec.~\ref{sec:update}. The update equations are put to use in a practical problem in Sec.~\ref{sec:experiment}, where an experiment uses two transformer classifiers on publicly available data. Sec.~\ref{sec:outlook} contains an outlook.

\section{A Dynamical Model For Classification}\label{sec:dynamical_model}
We introduce a dynamical model to temporally smooth class estimations. We start out with the simple case of a single classifier. After discussing state-evolution equations, we generalize that to multiple classifiers with different accuracies.

\subsection{Observation Model for a Single Classifier}
We assume that we observe $s(t) \in \mathbb{S}_n$, where $\mathbb{S}_n$ denotes the unit $n-1$ simplex. We interpret $s(t)$ as a sample from a Dirichlet distribution, parameterized by $\alpha(t) \in \mathbb{R}^n_+$. Here, $\alpha$ will act as the latent state we track over time. But to describe how we pose a state evolution equation, we first review the Dirichlet distribution along with a conjugate prior.

\subsection{Review of the Dirichlet Distribution}

Recall that a Dirichlet distribution, parameterized by  $\alpha \in \mathbb{R}^n_+$ (dropping the time index for simplicity) is given as \cite{bayesian, murphy2}
\begin{equation}
\text{Dir}(s |\alpha) = \underbrace{\dfrac{\Gamma\left(\sum_{i=1}^n\alpha_i\right)}{\prod_{i=1}^n\Gamma(\alpha_i)}}_{A(\alpha)}\,\exp\bigl(\langle \alpha - 1, \log(s)\rangle\bigr).
\end{equation}

A useful property of $A(\cdot)$ and $\Gamma(\cdot)$ is the following. 
\begin{prop}\label{prop:log-convex}
$\Gamma(\cdot):\mathbb{R}_+ \to \mathbb{R}$ is log-convex, $A(\cdot) : \mathbb{R}^n_+ \to \mathbb{R}$ is log-concave.
\end{prop}
These follow from the properties of exponential family (see, e.g., section 3.5 in \cite{wainwrightjordan})  - specifically, that the cumulant function is convex (and that $-\log \Gamma(\cdot)$ is the cumulant function for a specific gamma random variable).

Let us now see how to interpret the parameters of a conjugate prior for the Dirichlet distribution.
Suppose we observe $N$ independent vectors $s^1, \ldots, s^N$ sampled from this distribution (i.e., $s^i\in\mathbb{S}^n$). The joint probability distribution is of the form
\begin{multline}\label{eqn:Dirichlet}
p(s^1, s^2, \ldots, s^K | \alpha) \\= A(\alpha)^K\,\exp\left(\bigl\langle \alpha - 1, \sum_{k=1}^K \log \,s^k \bigr\rangle \right)
\end{multline}
This motivates the following distribution as a conjugate :
\begin{equation}\label{eqn:prior}
\text{CP}(\alpha | \eta, \nu) = A(\alpha)^{\eta}\,\exp\bigl(-\langle \alpha, \nu \rangle\bigr)\,B(\eta, \nu).
\end{equation}
This distribution was used in \cite{sta09p724} as a conjugate prior to the Dirichlet distribution. Comparing \eqref{eqn:prior} and \eqref{eqn:Dirichlet}, we can interpret $\eta \in \mathbb{R}_+$ as the number of observations, and $\nu \in \mathbb{R}^n$ as the sum of the logarithm of the inverse probability vectors.

The normalization constant $B(\eta,\nu)$ in  \eqref{eqn:prior} is not available  in closed form. However, we can compute the mode of \eqref{eqn:prior} with an iterative algorithm, as will be detailed in Sec.~\ref{sec:updating_posterior}, to sidestep the unavailability of $B(\eta,\nu)$.

\subsection{State Transition Model}
We propose a state-space model that respects the conjugacy relations, so that we end up with simple update steps. Given the whole past, the latent state $\alpha(t)$ is assumed to be distributed as
\begin{equation}\label{eqn:cur_alpha}
p(\alpha(t) | s(t), s(t-1), \ldots) = \text{CP}\bigl(\alpha(t) | \eta(t), \nu(t) \bigr).
\end{equation}
The parameters $\eta(t)$, $\nu(t)$, encode our knowledge of $\alpha(t)$. 

We postulate that in the absence of new observations, the next state follows the distribution
\begin{multline}\label{eqn:next_alpha}
p(\alpha(t+1) | \alpha(t), s(t), s(t-1), \ldots) \\= \text{CP}\bigl(\alpha(t+1) | \gamma \cdot \eta(t), \gamma \cdot v(t) \bigr),
\end{multline}
for $0<\gamma<1$.
It can be shown that 
\begin{equation*}
\text{CP}(\cdot | \gamma\cdot\eta, \gamma\cdot\nu) \propto \Bigl( \text{CP} (\cdot | \eta, \nu ) \Bigr)^{\gamma}.
\end{equation*} 
Raising the distribution $\text{CP}$ to a power $\gamma < 1$ preserves the mode, and increases the variance of the distribution. For a more detailed theoretical justification of this update rule, we refer to \cite{smi79p375}.

Finally, for the case of a single classifier, given $\alpha(t)$, we can postulate that the observations are assumed to be independent of the whole past, and are distributed as
\begin{equation}\label{eqn:obs}
p(s(t) | \alpha(t) ) \sim \text{Dir}(s(t) | \alpha(t)).
\end{equation}

For the single classifier use case, equations \eqref{eqn:cur_alpha}, \eqref{eqn:next_alpha}, \eqref{eqn:obs} determine the model that allow us to smooth the observed set of probability vectors and estimate $\alpha(t)$'s. Recall that, each $\alpha(t)$ is `summarized' by $\eta(t)$ and $\nu(t)$. In principle, all we need to do is to update $\eta(t)$ and $\nu(t)$. 

\subsection{Multiple Classifier Observation Model}\label{sec:multiple}

To handle multiple classifiers with varying accuracy, we introduce a mechanism to control the amount of uncertainty in the observations, without changing the state update equation postulated in \eqref{eqn:next_alpha}. 
To that end, we introduce a time-dependent constant $0 \leq \beta_t \leq 1$ and assume that given the underlying state, $\alpha(t)$, the observations $s(t)$ are distributed as 
\begin{equation*}
s(t) | \alpha(t)  \sim  \text{Dir}( \cdot | \beta_t \, \alpha(t) + (1- \beta_t)\,\mathbf{1}),
\end{equation*}
 where $\mathbf{1}$ denotes the vector of ones.  This is a heuristic choice, motivated by the fact that $\text{Dir}(\cdot | \mathbf{1})$ is the uniform distribution on the unit simplex. The idea is to use a high $\beta_t$ if the classification is reported by an accurate classifier, and a low $\beta$ otherwise. When we have multiple classifiers whose accuracies can be ordered, we can set $\beta_t$ according to this order.

\section{Update Equations}\label{sec:update}

We derive rules for updating $\eta(t)$ and $\nu(t)$ given a new observation $s(t)$.
We start by studying the posterior distribution for $\alpha$ given a new observation.

\subsection{Posterior Distribution}

The posterior distribution of $\alpha(t)$ given new class probability estimates $s(t)$ satisfies
\begin{multline*}
p(\alpha(t) | s(t), \nu(t-1), \eta(t-1)) \\\propto p(s(t) | \alpha(t))\cdot p(\alpha(t) | \nu(t-1), \eta(t-1)).
\end{multline*}
For simplicity, from this point on, we drop time indices and use $^\ast$ to refer to variables with time-index $t-1$.

Taking logarithms, we find,
\begin{multline}\label{eqn:logpost}
\log p(\alpha | s, \nu^*, \eta^* ) \propto \log A(\beta_t \alpha + (1-\beta)\,\mathbf{1}) \\+ \beta_t \langle \alpha, \log s \rangle + \gamma \eta\,\log A(\alpha) - \gamma \langle \alpha, \nu^*\rangle. 
\end{multline}

This distribution no longer belongs to the family \eqref{eqn:prior}. To approximate it  with a distribution from the family \eqref{eqn:prior}, we could resort to a variational framework and seek the closest member of the family with respect to the Kullback-Leibler divergence \cite{tzi08p131}. An alternative is to match moments as is common in assumed density filtering \cite{Sarkka_2013}. However, both approaches are infeasible due to the lack of closed form expressions when dealing with some of the terms in \eqref{eqn:logpost}. Instead, we consider the mode of the distribution for guiding the updates of $\eta$ and $\nu$. 

Working with the maximizer of \eqref{eqn:logpost} is feasible because the function is concave. This concavity property is a consequence of the log-concavity of $A(\cdot)$  (recall Prop.~\ref{prop:log-convex}), since  \eqref{eqn:logpost} is a linear combination of such functions and linear terms. Therefore, the mode is well defined. Armed with this observation, we first find the mode of the posterior distribution. We then choose $\mu$, $\eta$, following heuristic rules that will be detailed.

\subsection{Computing the Mode of the Posterior}
We expand  $\log A(\beta_t \alpha)$ to  write the rhs of \eqref{eqn:logpost} as
\begin{multline}\label{eqn:objective}
L(\alpha) := \log \Gamma\biggl(K\,(1-\beta_t) + \beta_t\, \sum_{i=1}^K \alpha_i \biggr) \\- \sum_{i=1}^K \log \Gamma\bigl(\beta_t\, \alpha_i + (1-\beta)\bigr) + \beta_t \, \langle \alpha, \log s \rangle \\+ \gamma\,\eta\,\left[ \log\Gamma\biggl(\sum_{i=1}^K \alpha_i \biggr) - \sum_{i=1}^K \log \Gamma(\alpha_i) \right]  - \gamma \langle \alpha, \nu \rangle ,
\end{multline}
where $K$ is the number of classes (or the dimension of $\alpha$).
Setting the gradient to zero to find the maximizer, we find that the mode must satisfy, for each $j=1,2,\ldots, K$,
\begin{multline}\label{eqn:optimality}
\beta_t \Psi\biggl(K\,(1-\beta_t) + \beta_t\, \sum_i \alpha_i \biggr) - \beta_t \Psi \left(\beta_t\, \alpha_j + (1-\beta)\right) \\+ \beta_t \log s_j + \gamma \eta \left[\Psi\biggl(\sum_i \alpha_i \biggr) - \Psi(\alpha_j) \right] - \gamma\,\nu_j = 0,
\end{multline}
where $\Psi(\cdot) = \partial \log \Gamma(\cdot)$ is the digamma function. Rearranging, we find that for each $j$, we have
\begin{multline*}
\beta_t \Psi (\beta_t\, \alpha_j + 1 - \beta_t) + \gamma \eta \Psi(\alpha_j) \\ = \beta_t \Psi\biggl( K\,(1-\beta_t) + \beta_t\, \sum_i \alpha_i \biggr) + \beta_t \log s_j \\+ \gamma \eta \Psi\biggl(\sum_i \alpha_i\biggr) - \gamma\,\nu_j.
\end{multline*}
These equalities are ripe for fixed point iterations of the form
\begin{multline*}
\beta_t \Psi (\beta_t\, \alpha^{k+1}_j + 1 - \beta_t) + \gamma \eta \Psi(\alpha^{k+1}_j) \\= \beta_t \Psi\biggl(K\,(1-\beta_t) + \beta_t\, \sum_i \alpha^{k}_i \biggr) + \beta_t \log s_j \\+ \gamma \eta \Psi\biggl(\sum_i \alpha^{k}_i\biggr) - \gamma\,\nu_j,
\end{multline*}
where the superscript in $\alpha^{k}$ denotes iteration index.
Let us  define a function
\begin{equation}
G_{\beta_t, c}(x) := \beta_t \Psi (\beta_t\,x + c) + \gamma \eta \Psi(x).
\end{equation}
Thanks to the strict monotonicity of $\Psi(\cdot)$, $G_{\beta_t, c}(x)$ is strictly increasing for positive $\beta_t$ and $c$. Therefore it has a well-defined inverse, allowing us to write 
\begin{equation}\label{eqn:iterations}
\alpha^{k+1}_i = G^{-1}_{\beta_t, (1-\beta_t)}\biggl( G_{\beta_t, K(1-\beta_t)}\Bigl(\sum_i\alpha_i^k\Bigr) + \beta_t \log y_i - \gamma\,\nu_i\biggr).
\end{equation}
In practice, we can invert $G_{\beta_t, c}(x)$ numerically with an algorithm like binary search \cite{Nocedal}.

Even though these iterations are well-defined, it is not obvious if they will converge. 
The following result establishes that the iterations are well-behaved in the sense that they increase the log-likelihood \eqref{eqn:objective} at each iteration. 
\begin{prop}
Let $\alpha_k$ be defined as in \eqref{eqn:iterations}. Then, for $L(\cdot)$ as in \eqref{eqn:objective},
\begin{equation*}
L(\alpha^k) \geq L(\alpha^{k-1}).
\end{equation*}
\begin{proof}
In a nutshell, we show that the algorithm is an instance of a minorization-maximization algorithm (see e.g., \cite{Lange, hun04p30,fig07p980} for detailed treatments - see also \cite{minka_dirichlet} for a similar approach in the context of the Dirichlet distribution).

As noted in Prop.~\ref{prop:log-convex}, $\Gamma(\cdot)$ is log-convex. Therefore, by a basic property of differentiable convex functions \cite{HiriartFund, Boyd},
\begin{equation}\label{eqn:ineq_gamma}
\log\,\Gamma(y) \geq \log \Gamma(x) + \langle y-x, \,\underbrace{\partial \log \Gamma(x)}_{\Psi(x)}\rangle
\end{equation}

We will apply this to the terms of $L(\cdot)$ that contain a sum, using $\alpha^k$ as the starting point. 
To simplify notation, let
\begin{align*}
u^k &= \sum_i \alpha^k_i,\\
x^k &= K\,(1-\beta_t) + \beta_t\,u^k.
\end{align*}
Now, using $x^k$, we write
\begin{multline*}
\log \Gamma\biggl(K\,(1-\beta_t) + \beta_t\, \sum_i \alpha_i \biggr) \\\geq  \log \Gamma(x^k) + \langle \alpha - \alpha^k, \Psi(x^k)\rangle
\end{multline*}
Similarly, using $u^k$, we can write
\begin{equation}
\log \Gamma\biggl(\sum_i \alpha_i \biggr) \geq \log \Gamma(u^k) + \langle \alpha - \alpha^k, \Psi(u^k)\rangle
\end{equation}
By replacing the terms in $L(\cdot)$ that contain a sum, we obtain the following function
\begin{multline}
 M(\alpha;\alpha^k) :=\langle \alpha - \alpha^k, \Psi(x^k)\rangle \\- \sum_i \log \Gamma\bigl(\beta_t\, \alpha_i + (1-\beta)\bigr) + \beta_t \, \langle \alpha, \log y \rangle \\+ \gamma\,\eta\,\left[\langle \alpha - \alpha^k, \Psi(s^k)\rangle  - \sum_i \log \Gamma(\alpha_i) \right]  - \gamma \langle \alpha, \nu \rangle.
\end{multline}
For a constant $c^k$ that is independent of $\alpha$, $M$ satisfies
\begin{itemize}
\itemsep0em 
\item $M(\alpha^k; \alpha^k) =  L(\alpha^k) + c^k$
\item $M(\alpha; \alpha^k) \leq  L(\alpha) + c^k$
\end{itemize}
Therefore, if $\alpha^* \in \arg\max_{\alpha} M(\alpha; \alpha^k)$, we have $L(\alpha^*) \geq M(\alpha^*;\alpha^k) - c^k \geq M(\alpha^k;\alpha^k) - c^k =  L(\alpha^k)$. But if we write the optimality conditions for maximizing $M(\alpha;\alpha^k)$, we find them to be equivalent to \eqref{eqn:optimality}. Thus follows the claim.
\end{proof}
\end{prop}

\subsection{Updating the Posterior Distribution Parameters}\label{sec:updating_posterior}

Consider now the prior distribution  \eqref{eqn:prior} over $\alpha$ with parameters $\eta$ and $\nu$. 
The mode of the distribution satisfies,
\begin{equation}\label{eqn:maximization_problem}
\alpha = \arg\max_{\alpha}\, \eta\,\log A(\alpha) + \langle \alpha, \nu \rangle.
\end{equation}
Setting the gradient to zero, we find that the parameters satisfy the following equations at the mode $\alpha^*$, for all $j$:
\begin{equation}\label{eqn:mode}
\Psi(\alpha^*_j) - \Psi\biggl(\sum_i\alpha^*_i\biggr) + \nu_j / \eta = 0.
\end{equation}
Arguably a reasonable choice is to find an ($\eta$, $\nu$) pair that is close to the previous values, and that also satisfies \eqref{eqn:mode}.
However, a simpler heuristic, that  seems to perform well is to consider $\beta_t$ as the `weight' of the observation. Since $\eta$ was found to be associated with the number of observations, we therefore argue in favor of updating $\eta$ as
\begin{equation}
\eta \gets \gamma_t\,\eta + \beta_t
\end{equation}
Given this value of $\eta$, we set $\nu_i$'s so that \eqref{eqn:mode} is satisfied.

The resulting pseudocode is shown in Algorithm~\ref{algo:EstPMF}.

\begin{algorithm}
\begin{algorithmic}[1]

\Require $\eta \in \mathbb{R}$, $\nu \in \mathbb{R}
^n$: parameters of the prior distribution $\eqref{eqn:prior}$ given all the observed probability estimates from the classifier up to the latest observation, $\gamma$:  `decay' parameter as in \eqref{eqn:next_alpha}, $K$:  number of classes, $\beta$: `weight' of the current observation,
$s\in \mathbb{R}^n$: probability estimate of the classifier, $\alpha^*$: mode of the prior distribution 
\Ensure $p_i$, probability estimate of the $i^{\text{th}}$ class, for $i=1,2,\ldots, K$ given the latest observation
\For {$M$ iterations}
\State $\alpha \gets \alpha^*$
\For {$i\in\{1,2,\ldots, K\}$}
\State $r_i \gets G_{\beta, K(1-\beta)}\Bigl(\sum_i\alpha_i\Bigr) + \beta \log s_i - \gamma\,\nu_i$
\State $\alpha^*_i \gets G^{-1}_{\beta, (1-\beta)}( r_i)$ for $i=1,2,\ldots, n$.
\EndFor
\EndFor
\State $\eta \gets \gamma\,\eta + \beta$
\State $\nu_i \gets \eta\,\Bigl(\Psi(\alpha^*_i) - \Psi\bigl(\sum_i\alpha^*_i\bigr) \Bigr)$ for $i=1,2,\ldots, K$.

\State $p \gets  \text{Mode of  }\text{Dir}(\cdot | \alpha^*)$
\end{algorithmic}
\caption{Probability Estimate Updates\label{algo:EstPMF}}
\end{algorithm}

\section{Experiments}\label{sec:experiment}

To assess the usefulness of the proposed method, we conducted experiments on the Capture-24 dataset \cite{ger20p318}. A  subset of this dataset \cite{wil18} contains 3-axis accelerometer data in normal living conditions from 150 subjects, where subjects' activities are classified into six different classes : `cycling', `mixed', `sit-stand', `sleep', `vehicle', `walking'.

We need `weak' and `strong' base classifiers to demonstrate how the proposed filter operates. Aligned with the general theme of the paper, the performance of the two classifiers should vary and depend on their computational complexity. We chose transformers as the base model. We specifically altered the structure proposed in \cite{sha21p540} for our task. Our `weak' and `strong' transformers have $\sim$45K and $\sim$180K weights, and operate on 5 and 30 seconds of input, respectively. We forgo further details about their architecture and refer to \cite{sha21p540}, as our goal was to evaluate the filter using a modern approach with fairly good performance. We do think there can be room for improvement for these base classifiers, but investigating that is beyond the scope of the current paper.

We trained the base classifiers using data from 125 subjects, and tested on data from the remaining 25 subjects. All of the results reported below are from the test set.

Given the classifiers, we need a policy to decide when to use the output of which classifier. While this is an interesting decision problem in itself, we opted for a simple policy. We call the strong classifier every minute. In between, we resort to the weak classifier every 5 seconds.

We compared classification accuracy of four methods :
\begin{itemize}
\itemsep0em 
\item ``\textbf{Raw}": Classifier outputs without  smoothing
\item ``\textbf{Simple}'' : Running-average of class probabilities using a window of length $W=5$ (so the average spans a collection of class estimates obtained over 25 seconds)
\item ``\textbf{Single}'' : Proposed method with $\beta_t=1$ for all $t$ (i.e., not differentiating between weak and strong classifiers).
\item ``\textbf{Multiple}'' : Proposed method with $\beta_t = 1$, $\beta_t = 0.5$ for the strong and weak classifier respectively.
\end{itemize}

Percentages of correct classification are tabulated in Table~\ref{table:correct}. We see a clear trend. Smoothing even with a running average improves results. Using the proposed smoothing classifier without differentiating between weak and strong classifiers helps further boost results. That suggests that modeling the underlying distribution as a sample from a Dirichlet distribution has merit. Finally, treating weak and strong classifiers differently provides an additional boost.
\begin{table}[h]
\centering
\begin{tabular}{ c c c c}
\thickhline
Raw & Simple & Single  & Multiple \\
\hline
77.15 & 80.85 & 82.61 & 85.24 \\
\thickhline
\\
\end{tabular}
\caption{Percentage of Correct Classification}
\label{table:correct}
\end{table}

To get a better sense, we also computed sensitivity (probability of detection, or true positive rate), and specificity (true negative rate) for each class.
The results are tabulated in Tables~\ref{table:sensitivity}, \ref{table:specificity} respectively. Although the results are more nuanced, the outlined trends  are still valid. Some of these numbers are also affected by imbalance between the different classes.

\begin{table}
\centering
\begin{tabular}{c c c c c}
\thickhline
&Raw&Simple&Single&Multiple\\
\hline
bicycling&0.86&0.89&0.88&0.82\\
mixed&0.68&0.73&0.76&0.77\\
sit-stand&0.85&0.90&0.93&0.94\\
sleep&0.81&0.83&0.84&0.87\\
vehicle&0.65&0.72&0.75&0.81\\
walking&0.33&0.34&0.34&0.44\\
\thickhline
\\
\end{tabular}
\caption{Sensitivity : TP / (TP + FN)}
\label{table:sensitivity}
\end{table}

\begin{table}
\centering
\begin{tabular}{c c c c c}
\thickhline
&Raw&Simple&Single&Multiple\\
\hline
bicycling&0.79&0.84&0.87&0.88\\
mixed&0.60&0.66&0.69&0.75\\
sit-stand&0.73&0.76&0.78&0.80\\
sleep&0.93&0.95&0.97&0.98\\
vehicle&0.78&0.87&0.90&0.89\\
walking&0.60&0.71&0.75&0.73\\
\thickhline
\\
\end{tabular}
\caption{Specificity : TN / (TN + FP)}
\label{table:specificity}
\end{table}

\begin{figure}
\centering
\includegraphics[scale =0.4]{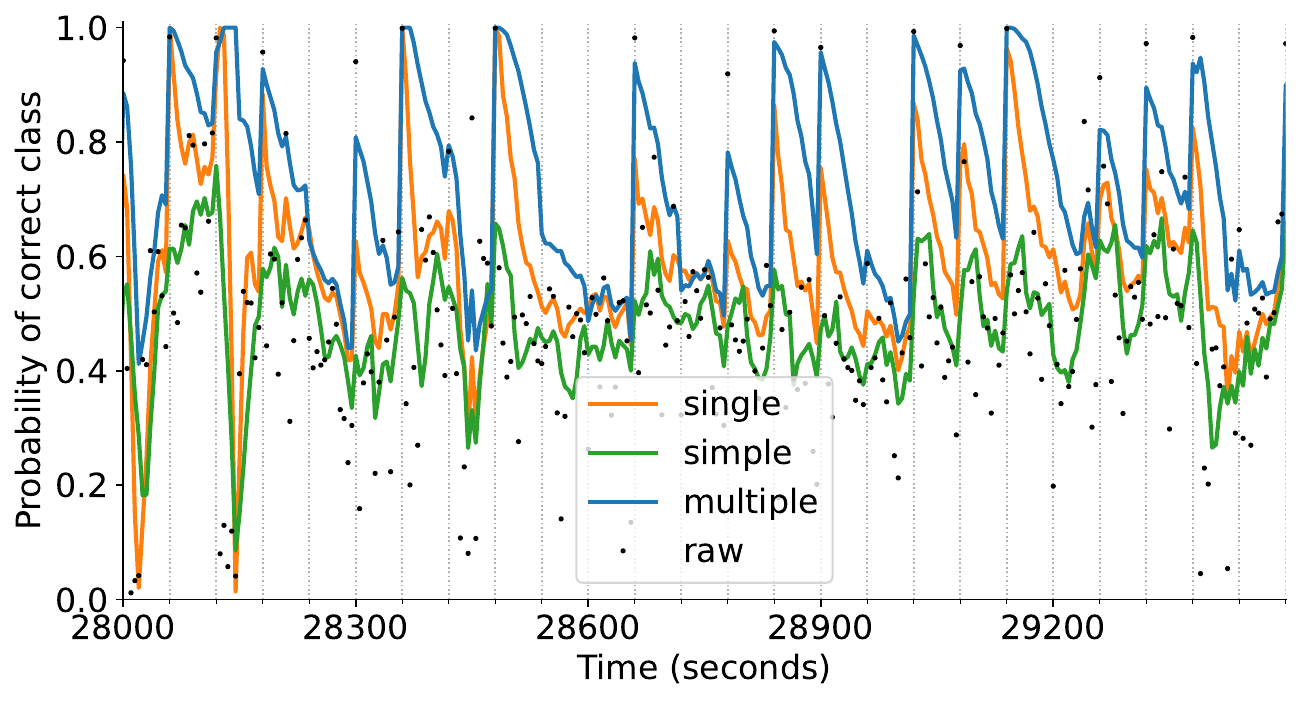}
\caption{Detailed excerpt from one of the test examples, showing the probability estimate for the true class. See text for a description. \label{fig:detail}}
\end{figure}

Finally, we consider a specific example to show how the proposed filter behaves in a desirable way.  Fig.~\ref{fig:detail} contains an excerpt from one of the test subjects. 
All of the probabilities shown in this graph belong to the true class - i.e., ideally we would see all tracks equal to unity.
Black dots indicate the raw probability obtained from the base (transformer) classifiers. Calls to the strong classifier are marked with vertical grid lines. We see that in this snippet, the strong classifier is performing much better than the weak classifier as its output is considerably larger. The `simple' running average computes the average of the black points. We do observe that averaging reacts to the occasional better estimates from the strong classifier. Similarly, the `single' smoother, that treats the weak and strong classifiers equally is influenced by the strong classifier. However, the proposed method (`multiple') that treats strong and weak classifiers differently exhibits a more desirable pattern. The high probability for the true class is held for a longer period, resulting in improved accuracy. 

\section{Outlook}\label{sec:outlook}
One of the aspects that we did not discuss is practical implementation. Evaluating a special function like the digamma function on a battery-powered device is not desirable. Instead, one can resort to look-up tables or simple approximations as in \cite{minka_dirichlet}. In our experience from different contexts, this does not degrade performance noticeably.

We also left the determination of the classifier noise level, $\beta$, to the user. However, if the user already has an idea about the confusion matrix associated with each classifier, those can, in principle, be used to set $\beta$. 

Finally, finding a good selection policy given classifiers is potentially a deeper problem, solution of which can directly help boost accuracy.

\end{document}